\documentclass[manuscript,nonacm]{acmart}
\usepackage{amsmath}
\usepackage{hyperref}
\usepackage{cleveref}
\usepackage[figuresright]{rotating}
\newcommand{\youtube}{YouTube}
\usepackage{ulem}
\usepackage{float}
\usepackage{adjustbox}
\usepackage{standalone}
\usepackage{tikz}
\usetikzlibrary{arrows,shapes,positioning, matrix}
\usepackage[inline]{enumitem}
\usepackage{booktabs}
\usepackage{pifont}
\usepackage{textcase}
\usepackage{subcaption}
\usepackage{multirow}
\usepackage{wasysym}
\usepackage{fontawesome}

\definecolor{mforestgreen}{HTML}{228B22}

\definecolor{mforestgreen}{HTML}{228B22}
\usepackage{ifthen}
\usepackage{ulem}
\newboolean{mocommentenabled}
\newboolean{onlysuggestedtext}

\newcommand{\bftab}{\fontseries{b}\selectfont}
\definecolor{mforestgreen}{HTML}{228B22}

\begin{document}

\title{Embedding-Based Rankings of Educational Resources based on Learning Outcome Alignment: Benchmarking, Expert Validation, and Learner Performance}

\author{Mohammadreza Molavi}
\email{Mohammadreza.molavi@tib.eu}
\affiliation{%
  \institution{Leibniz Information Centre for Science and Technology (TIB)}
  \country{Germany}
}

\author{Mohammad Moein}
\email{mohammad.moein@tib.eu}
\affiliation{%
  \institution{Leibniz Information Centre for Science and Technology (TIB)}
  \country{Germany}
}

\author{Mohammadreza Tavakoli}
\email{reza.tavakoli@tib.eu}
\affiliation{%
  \institution{Leibniz Information Centre for Science and Technology (TIB)}
  \country{Germany}
}

\author{Abdolali Faraji}
\email{abdolali.faraji@tib.eu}
\affiliation{%
  \institution{Leibniz Information Centre for Science and Technology (TIB)}
  \country{Germany}
}

\author{Stefan T. Mol}
\email{s.t.mol@uva.nl}
\affiliation{%
  \institution{University of Amsterdam}
  \country{Netherlands}
}

\author{G\'abor Kismih\'ok}
\email{Gabor.Kismihok@tib.eu}
\affiliation{%
  \institution{Leibniz Information Centre for Science and Technology (TIB)}
  \country{Germany}
}

\renewcommand{\shortauthors}{Molavi et al.}

\begin{abstract}
As the online learning landscape evolves, the need for personalization is increasingly evident. Although educational resources are burgeoning, educators face challenges selecting materials that both align with intended learning outcomes and address diverse learner needs. Large Language Models (LLMs) are attracting growing interest for their potential to create learning resources that better support personalization, but verifying coverage of intended outcomes still requires human alignment review, which is costly and limits scalability. We propose a framework that supports the cost-effective automation of evaluating alignment between educational resources and intended learning outcomes. Using human-generated materials, we benchmarked LLM-based text-embedding models and found that the most accurate model (\textit{Voyage}) achieved 79\% accuracy in detecting alignment. We then applied the optimal model to LLM-generated resources and, via expert evaluation, confirmed that it reliably assessed correspondence to intended outcomes (83\% accuracy). Finally, in a three-group experiment with 360 learners, higher alignment scores were positively related to greater learning performance, $\chi^{2}(2, N = 360) = 15.39$, $p < .001$. These findings show that embedding-based alignment scores can facilitate scalable personalization by confirming alignment with learning outcomes, which allows teachers to focus on tailoring content to diverse learner needs.

\end{abstract}

\maketitle

\section{Introduction and Background}\label{sec1}
Online education has expanded markedly in recent years, driven by learners’ routine use of online platforms, the COVID-19 pandemic, and a renewed emphasis on lifelong learning, positioning digital tools as critical for equity and inclusiveness \cite{Greenhow_Graham_Koehler_2022, Zhang_Carter_Qian_Yang_Rujimora_Wen_2022, tavakoli2022ai}. In this online landscape, the demand for personalization and inclusiveness underscores the challenge of curating content that aligns with both target learning outcomes and diverse learner needs \cite{denny2023can, tavakoli2022ai}. Although many resources are available, teachers find the process of identifying suitable materials time-consuming and inefficient \cite{molavi2020extracting, deldjoo2020recommender}. Moreover, prioritizing personalization can inadvertently weaken alignment, necessitating deliberate safeguards \cite{latent2025curriculardrift}. Prior work has sought technological remedies: early systems aggregated open educational resource repositories \cite{mason, mourino2018cross} but did not aid selection because they lacked ranking sensitive to pedagogical context; subsequent approaches used semantic technologies and knowledge graphs to encode teaching context \cite{estivill2019towards, Limongelli2022}, yet they face scalability limits due to high development and maintenance costs \cite{wang2020semantic}. More scalable machine-learning techniques—such as \textit{Learning to Rank (LTR)} and topic modeling—have also been explored \cite{usta2021learning, Hariharan2011}, but accuracy remains subpar.

Teachers typically engage in three key tasks: aligning instructional content with intended learning outcomes, delivering instruction, and personalizing learning experiences \cite{bishop2020teacher}. Of these, alignment is particularly amenable to automation, whereas effective teaching and personalization depend on students’ contexts and therefore require teachers’ nuanced judgment and interaction \cite{bishop2020teacher}. Large Language Models (LLMs), with their advanced natural language processing and reasoning capabilities \cite{zhu2023large}, offer a promising new direction. They can potentially support teachers by efficiently identifying those resources that are constructively aligned \cite{biggs1996enhancing} with learning outcomes. Furthermore, LLMs can generate new educational materials, expanding the range of resources available to educators \cite{denny2023can} and effectively reducing their workload.

Despite the promise of LLMs, incorporating LLMs in education still presents significant challenges. To date, generated educational content requires careful verification to ensure alignment with learning outcomes \cite{westerlund2024llm, biggs1996enhancing}, and the high computational costs of these models limit their accessibility and scalability \cite{wang2024large}. While recent research has explored using LLMs to create content in narrowly defined domains such as programming \cite{denny2023can, westerlund2024llm, jury2024evaluating}, these efforts highlight the persistent risks of hallucination and the high costs associated with quality control \cite{abu2024knowledge, christensen2023llm}.

In this paper, we explore cost-effective and scalable LLM-based techniques to support teachers by providing resource rankings. These rankings, that are based on the alignment of a resource with intended learning outcomes, can be applied to evaluate either existing or generated educational content. By relying on these rankings, teachers can ensure their pedagogical goals are met, while simultaneously allowing them to tailor resources to diverse learner needs, such as accessibility requirements or varying levels of prior knowledge. To achieve this, we set out to answer the following research questions:

\begin{enumerate}[label=(\arabic*)]
    \item Can text embeddings effectively expose the alignment between a candidate educational resource and teachers' intended learning outcomes?
    \item If so, can embedding-based alignment rankings of LLM-generated resources be validated through expert evaluation and, subsequently, shown to predict improved learning performance?
\end{enumerate}

\begin{figure}[t]
    \centering
        \includegraphics[width=\textwidth]{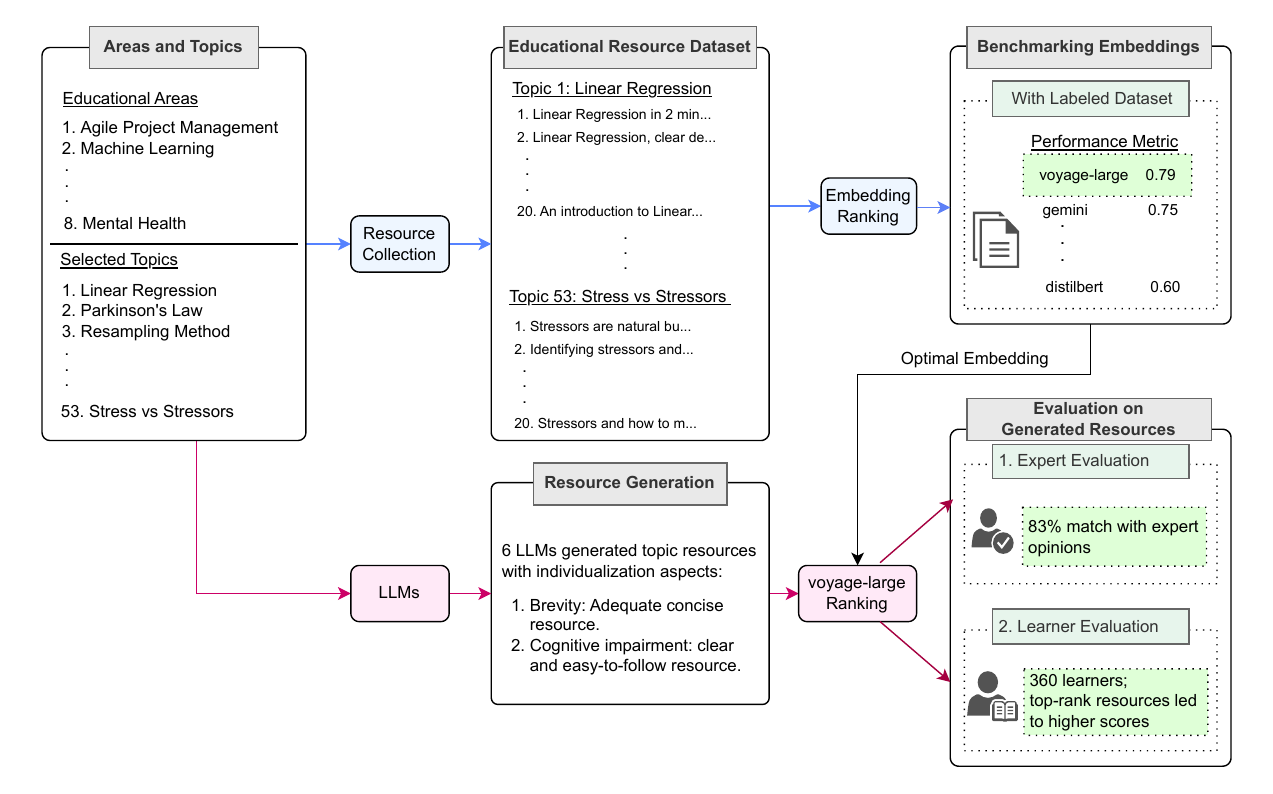}
    \vspace{-0.7cm}    
    \caption{Our methodology has two main components: (1) benchmarking how text-embedding models rank resources based on alignment with learning outcomes using our collected labeled dataset (\textcolor{blue}{blue flow}), and (2) assessing the alignment-ranking quality of the optimal embedding model on LLM-generated resources—via expert and learner evaluations (\textcolor{purple}{red flow}).}
    \label{fig:toc}
\end{figure}

To answer these questions, we conducted two studies. The first study evaluated how effectively various text embedding models—from prominent models such as \textit{Google Gemini} and \textit{OpenAI ChatGPT} to open-source alternatives \cite{muennighoff-etal-2023-mteb}—ranked existing resources from \youtube{} against intended learning outcomes. We developed a scoring metric inspired by Kendall’s tau \cite{kendall1938} to assess ranking quality, prioritizing models that ranked resources aligned with intended learning outcomes above those not aligned. The second study exploited the best-performing model from the first study to rank LLM-generated resources in an effort to examine whether it would also be deemed superior by experts and yield enhanced learning performance. Here, six different LLMs were prompted to generate educational content tailored to specific personalization and inclusiveness use cases. These generated resources were again labeled by experts, and we used our ranking score to assess the ability of the optimal model to evaluate them. As an extension, we further tested whether these rankings are associated with superior learning performance. Using an experimental design with 360 participants, we found that higher-ranked resources consistently resulted in better performance outcomes, demonstrating that our ranking approach not only aligns with expert judgment but also predicts learning performance.

The findings from this three-level evaluation demonstrate that text-embedding-based ranking methods can effectively support teachers in discovering and generating personalized, inclusive resources that align with their intended learning outcomes. Specifically, our approach assesses this alignment so that teachers are able to focus on learners’ idiosyncratic needs rather than merely checking for constructive alignment between learning resources and learning objectives. \Cref{fig:toc} summarizes the pipelines of the conducted studies.

\section{Methods and Results}\label{sec:methods}
In this section, we present a framework\footnote{The source code can be accessed at \href{https://2ly.link/27a3O}{\texttt{https://2ly.link/27a3O}}} for studying LLM-based ranking of educational resources with respect to teachers' target learning outcomes. We evaluated several LLM-based text embedding models to assess the alignment between resources and intended learning outcomes, offering a cost-efficient way to leverage the deep semantic capabilities of LLMs without the computational burden of full-text generation. In addition, we examined how the optimal ranking technique could help teachers evaluate generated resources tailored to different learners’ needs. Finally, to test whether these rankings translate into measurable learning benefits, we conducted a learner study in which participants used the ranked resources, and we measured their performance.

\subsection{Framework for Evaluating Resource Ranking Performance}

\subsubsection{Ground Truth Creation}
\paragraph*{Topic Selection} Following the concept from successfully completed and positively evaluated \textit{European Union} funded educational projects\footnote{The project names will be added after the blind reviews}, we decided to focus on the following educational domains: \textit{Agile Project Management}, \textit{Machine Learning}, \textit{Prompt Engineering}, \textit{Time Management}, \textit{Python Programming}, \textit{Mental Health}, \textit{First Aid Training}, and \textit{Using LaTex}. Based on the objectives of these projects, we compiled a diverse list of topics we wanted to focus on, which resulted in 53 separate topic titles.

\paragraph*{Educational Resource Gathering} We decided to use \youtube{} to collect educational resources,  since
\noindent\begin{enumerate*}[label=(\arabic*)]
    \item it is a major repository of learning resources that are widely used by informal/lifelong learners \cite{Pires_Masanet_Tomasena_Scolari_2022,Bello_Bravo_Payumo_Pittendrigh_2021}
    \item it provides a broad range of resources from different contexts (e.g., location, educational domains), and
    \item it offers a powerful ranking for search results, which we could benefit from as a baseline ranking algorithm.
\end{enumerate*}
Topic titles were entered as search queries on \youtube{}, and we retrieved the top 20 video results for each search query (provided that they were in English and included subtitles). These sets of videos served as the baseline ranking of the resources for the topics. Although we opted to focus on \youtube{} for the purposes of this paper, it is important to note that because our model relies on resource transcripts, it is  broadly applicable to any type of educational material that is either text-based or can be converted to text.

\paragraph*{Data Labeling}  The collected resources were classified by domain experts—two per topic, each with at least five years of relevant teaching experience—into two main categories: \textit{'accepted'} and \textit{'rejected'}. \textit{Accepted} resources were those considered to be constructively aligned with the intended learning outcomes, while \textit{rejected} resources consisted of those deemed either irrelevant, those not covering all the learning outcomes, and/or those covering irrelevant topics. At the end of this step, the ground truth was established as $D=\{T_1, \ldots , T_{53}\}$ where $T_i$ represents a topic\footnote{Ground truth data in SQLite3: \href{https://2ly.link/27QHn}{\texttt{https://2ly.link/27QHn}}.}. Each topic $T_i$ was linked to a set of resources and their concomitant transcripts $R_{ij}$. Additionally, there was a set of labels $L_i=\{l_{i1}, \ldots, l_{ij}\}$, associated with $T_i$, where $l_{ij}$ was the label (either \textit{accepted} or \textit{rejected}) for the resource $R_{ij}$.

\subsubsection{Ranking Educational Resources}
\paragraph*{Embedding-Based Ranking} Since identifying teachers' intended learning outcomes was not always feasible, we used the transcripts of \textit{accepted} resources as proxies for their intended outcomes. This aligns with real-world scenarios where a teacher locates a suitable resource and then seeks similar materials to support diverse learners' needs. To rank resources within each topic, we leveraged text embedding models. In each step, we selected an \textit{accepted} resource as the target learning outcome and ranked other resources based on their alignment with the selected resource, which we operationalized in terms of the \textit{Cosine similarity} between embedding vectors. Specifically, we opted for efficient and widely adopted text embeddings \cite{jing2024large}, using the following language models: \textit{OpenAI/text-embedding-3-small}, \textit{OpenAI/text-embedding-3-large}, \textit{OpenAI/text-embedding-ada-002}, \textit{Voyage/voyage-large-2-instruct}, \textit{Google\allowbreak/gemini-1.0}, \textit{BAAI/bge-large-en-v1.5}, and \textit{BAAI/bge-small-en-v1.5}. These embeddings exhibit superior performance in handling long-context dependencies and in understanding context compared to traditional embeddings \cite{Keraghel2024BeyondWA,Freestone2024WordER}. However, since these models impose a maximum length limit and truncate text exceeding it, we segmented each transcript into smaller parts within the allowed limit. Embeddings were computed separately for each segment and then averaged to generate a single vector representation of the complete transcript. Additionally, as a candidate from traditional embedding models, we also evaluated DistilBERT (specifically, \textit{distilbert-base-uncased}) to provide a comparative embedding-based baseline.

\vspace{-0.1cm}
\begin{figure}[t]
    \centering
    \resizebox{\textwidth}{!}{
        \includegraphics{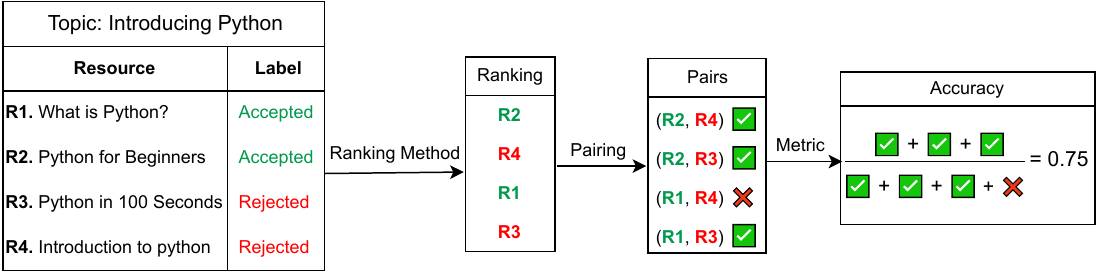}
    }
    \vspace{-0.5cm}
    \caption{Sample calculation of accuracy for a topic with four resources. Accuracy was determined by the proportion of correctly ranked accepted-rejected pairs relative to all such pairs.}
    \label{fig:metricpipeline}
\end{figure}

\begin{equation}
    \label{eq:accuracy}
    \begin{aligned}
        accuracy(O_i) & = \frac{count(R_{ij}, R_{ik}) \text{ if }R_{ij}\text{ is ranked higher than }R_{ik} \text{ in }O_i}{count(R_{ij}, R_{ik})} \\
        O_i           & = \text{a ranking for topic }T_i                                                                                           \\
        R_{ij}        & = \text{a resource in } O_i \text{ where }l_{ij} = \textit{accepted}                                                       \\
        R_{ik}        & = \text{a resource in } O_i \text{ where }l_{ik} = \textit{rejected}                                                       \\
    \end{aligned}
\end{equation}

\paragraph*{Ranking Evaluation Metric} The evaluation objective was to compare the generated rankings to identify the most effective model. Therefore, we needed to define a metric that prioritized the rankings that put \textit{accepted} resources in higher positions in the list (and accordingly the \textit{rejected} resources in lower positions). We propose that the ideal ranking would be a ranking in which for each pair of $(R_{ij}, R_{ik})$ of topic $T_i$ where $l_{ij} = \textit{accepted}$ and $l_{ik} = \textit{rejected}$, $R_{ij}$ appears in a higher position than $R_{ik}$. We defined our metric to consider how accurately a ranking can resolve such pairs of $(R_{ij}, R_{ik})$. The more pairs the ranking can resolve, the higher the accuracy score. This approach is inspired by Kendall’s tau \cite{kendall1938}, a classical measure of rank correlation that evaluates the degree of agreement between pairwise orderings, while adapting its core premise to better suit our specific evaluation needs. \Cref{fig:metricpipeline} illustrates a sample calculation, and equation \ref{eq:accuracy} defines the $accuracy$ function. Additionally, since in many practical scenarios users tend to examine only the top of the ranked list, we computed Precision@k \cite{manning2009introduction} as a complementary metric to assess the precision of top-ranked resources, specifically using \(k=3\) and \(k=5\).

\begin{table}[h]
    \centering
    \begin{adjustbox}{max width=\textwidth,max height=0.5\textheight }

\begin{tabular}{ll}
	\toprule
	                             Model  & Accuracy $\pm$ SD   \\
	\midrule
	\textbf{Voyage/voyage-large-2-instruct} & \textbf{0.79 ± 0.14} \\
	Google/gemini-1.0              & 0.75 ± 0.15 \\
	OpenAI/text-embedding-3-large  & 0.73 ± 0.15 \\
	BAAI/bge-small-en-v1.5         & 0.71 ± 0.17 \\
	OpenAI/text-embedding-3-small  & 0.69 ± 0.16 \\
	OpenAI/text-embedding-ada-002  & 0.68 ± 0.15 \\
	BAAI/bge-large-en-v1.5         & 0.65 ± 0.18 \\
	DistilBERT/distilbert-base-uncased        & 0.60 ± 0.16 \\
	Baseline (YouTube)       & 0.59 ± 0.18 \\
	\bottomrule
\end{tabular}

    \end{adjustbox}
    \caption{Average ranking accuracy of the baseline and different models.}
    \label{tab:results}
\end{table}

\vspace{-0.7cm}
\subsubsection{Ranking Results and Discussion}
\begin{figure*}[t]
    \centering
    \includegraphics[width=\textwidth]{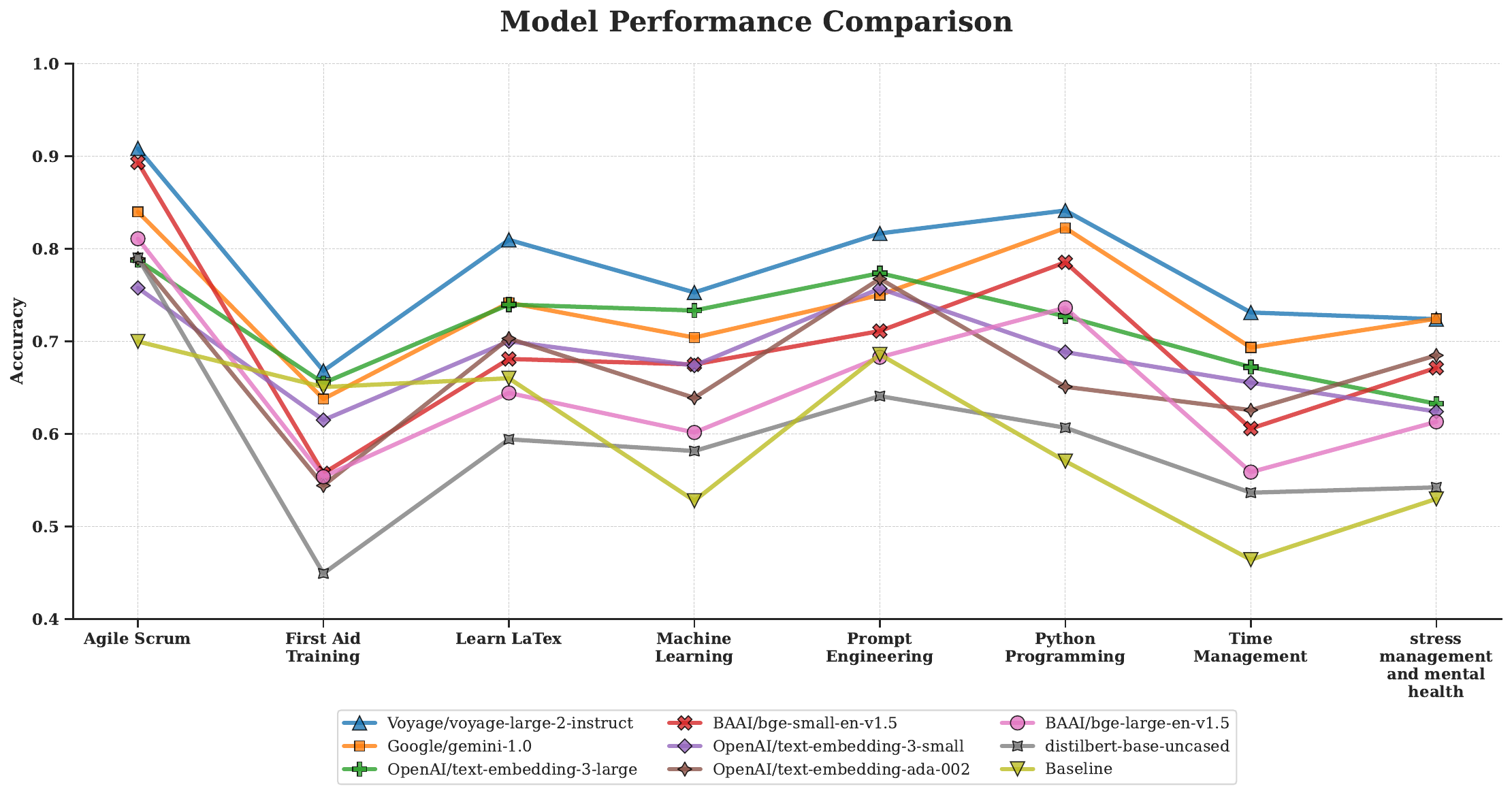}
    \vspace{-0.4cm}
    \caption{Average ranking accuracy per domain for each model.}
    \label{fig:area_perf}
\end{figure*}

\paragraph{Results} \Cref{tab:results} presents the average ranking accuracy for each model\footnote{Ranking accuracy per topic for different embedding models: \href{https://2ly.link/27QHF}{\texttt{https://2ly.link/27QHF}.}}. The Voyage embedding model outperformed all other models. Furthermore, we analyzed the average performance across the specified educational domains and observed a consistent ranking, as shown in \mbox{\Cref{fig:area_perf}}.
A Friedman test comparing the 8 models and the baseline showed a clear difference in performance \((\chi^2(8)=142.65,\ p<.001)\), with a medium effect size (Kendall’s W=.34). Nemenyi post-hoc comparisons revealed that the \textit{Voyage} model significantly outperformed every other model (after correction for type 1 errors), except \textit{Gemini}, where the difference favored Voyage but was not statistically significant. No other model surpassed Voyage in any pairwise comparison. This analysis was performed across 53 topics. Additionally, the Precision@3 and Precision@5 results presented in \Cref{tab:p@k} further confirm that the Voyage embedding model outperformed the other models in retrieving relevant resources at the top of the ranking\footnote{Precision@k per topic for different embedding models: \href{https://2ly.link/27QHH}{\texttt{https://2ly.link/27QHH}.}}.

\begin{table}[t]
    \centering
    \begin{adjustbox}{max width=\textwidth, max height=\textheight}
   {

\begin{tabular}{lll}
    \toprule
    Model                         & Precision@3 & Precision@5 \\
    \midrule
    \bftab {Voyage/voyage-large-2-instruct} & \bftab{0.68 ± 0.25} & \bftab{0.61 ± 0.25} \\
    OpenAI/text-embedding-3-large  & 0.62 ± 0.29 & 0.57 ± 0.26 \\
    Google/gemini-1.0              & 0.60 ± 0.28 & 0.55 ± 0.25 \\
    BAAI/bge-small-en-v1.5         & 0.57 ± 0.29 & 0.53 ± 0.26 \\
    OpenAI/text-embedding-3-small  & 0.57 ± 0.28 & 0.51 ± 0.25 \\
    Baseline                       & 0.55 ± 0.31 & 0.51 ± 0.26 \\
    OpenAI/text-embedding-ada-002  & 0.55 ± 0.28 & 0.50 ± 0.25 \\
    BAAI/bge-large-en-v1.5         & 0.53 ± 0.29 & 0.48 ± 0.24 \\
    distilbert-base-uncased        & 0.44 ± 0.25 & 0.44 ± 0.22 \\
    \bottomrule
\end{tabular}
}
\end{adjustbox}
    \caption{Precision@3 and Precision@5 scores for all models.}
    \label{tab:p@k}
\end{table}

\paragraph*{Discussion} The results showed that the ranking methods can viably automate the evaluation of the quality of educational resources when it comes to the alignment with the intended learning outcomes, which can support teachers in gaining efficient access to available high-quality resources. This claim is evidenced by the fact that the optimal model (i.e., Voyage) obtained better ranking and precision scores than the baseline model, which was the \youtube{} ranking. \youtube{}, as a major learning platform, benefits from a comprehensive set of features to rank the videos \cite{rieder2018ranking,crack}. Therefore, outperforming \youtube{} shows the effectiveness of our top-ranking models. Moreover, the open-source \textit{bge-small} model also performed reasonably well with an accuracy of $0.71$, meaning it can be a competitive choice in cases where privacy, copyright infringement,  and/or cost are a concern.

\subsection{Evaluating the Optimal Ranking of Generated Resources by Experts and Learners}
\paragraph*{Objective}
Utilizing our ranking framework, we investigated our second research question: whether embedding-based alignment rankings of LLM-generated resources can be validated through expert evaluation and, subsequently, shown to predict improved learning performance.  To evaluate a personalization-oriented use case, we focused on the brevity (level of detail) of the generated resources, following \cite{tavakoli2022ai}. Indeed, brevity is one key feature on which learning resources may be personalized, as some learners may desire or  require greater detail than others, who may prefer not being burdened by excessive detail. As the second use case, we examined inclusiveness, specifically evaluating accessibility for individuals with cognitive impairments following \cite{elias2017ontology}. For our analysis, we employed six leading large language models: three \textit{OpenAI ChatGPT} models (\textit{GPT-4o}, \textit{GPT-4o-mini}, and \textit{GPT-3.5}) and three \textit{Google Gemini} models (\textit{gemini-1.5-flash, gemini-1.5-pro,} and \textit{gemini-2-flash}). These models were selected based on their prominence in the field \cite{gpt} and their availability via APIs, which facilitates replicability of our approach and findings. To ensure an accurate assessment of the usefulness of generated educational content, we focused on two of our target educational domains: Python programming and Machine learning. These domains encompassed a total of 22 topics. It is important to note that this study specifically evaluated how closely rankings from our optimal solution (Voyage) align with human expert choices regarding the usefulness of the content in covering all intended learning outcomes while accommodating diverse learners’ needs.

\paragraph*{Resource Generation}
The same two domain experts per topic from the first study were asked to specify the learning outcomes they would cover for each of the 22 topics, according to one of the videos they accepted in the first study. Next, we prompted six LLMs to generate resources for each topic that addressed the intended learning outcomes. Each LLM was instructed to generate two types of resources per topic: (1) \textit{brief} resources with concise and to-the-point content, designed for learners who prefer a quick, less detailed approach \cite{tavakoli2022ai}, and (2) resources tailored to individuals with cognitive impairments, featuring step-by-step explanations, clear language, short sentences, and practical examples. These features were designed in collaboration with a specialist in cognitive accessibility and guided by principles derived from the relevant literature \cite{friedman2007web}. In sum, in total, 264 resources (6 LLMs * 22 topics * 2 use cases) were generated.

\paragraph*{Evaluation on the Generated Resources}
In the generation step, six resources were generated for each topic and for each individualization use case (i.e., personalization and inclusiveness). The two domain experts collaboratively evaluated these resources through consensus building. They only classified a resource as accepted if they both found it useful and suitable for the topic at hand. The criteria for acceptance included, but were not limited to: covering all intended learning outcomes, containing no incorrect or inaccurate information, and avoiding the inclusion of unrelated or extraneous topics. Next, we ranked the resources using the optimal solution (i.e., Voyage) and cosine similarity to a reference resource (an accepted resource randomly selected from the first study). Similar to the first study, our ranking evaluation score was calculated by analyzing all accepted-rejected  pairs and determining the proportion of cases where accepted resources received a higher rank than rejected ones. The results indicated that our optimal solution achieved an average ranking accuracy of 0.83. \Cref{tab:individualization} presents the detailed findings of our study.

\begin{table}[h]
\setlength{\tabcolsep}{10pt}
    \centering
    \begin{adjustbox}{max width=\textwidth}
    
    \begin{tabular}{l l c c c}
        \toprule
        \multicolumn{1}{l}{\textbf{Individualization}} & \multicolumn{1}{l}{\textbf{Educational}} & \multicolumn{1}{l}{\textbf{Generated}} & \multicolumn{1}{l}{\textbf{Accepted}} & \multicolumn{1}{l}{\textbf{Ranking}} \\
        \multicolumn{1}{l}{\textbf{Aspect}} & \multicolumn{1}{l}{\textbf{Domain}} & \multicolumn{1}{l}{\textbf{Resources}} & \multicolumn{1}{l}{\textbf{Resources}} & \multicolumn{1}{l}{\textbf{Accuracy}} \\
        \midrule
        Brevity & Python Programming & 60 & 27 & 0.85 \\
        Brevity & Machine Learning   & 72 & 32 & 0.80 \\
        Cognitive impairment & Python Programming & 60 & 27 & 0.80 \\
        Cognitive impairment & Machine Learning   & 72 & 39 & 0.83 \\
        \bottomrule
    \end{tabular}
    \end{adjustbox}
    
    \caption{Ranking evaluation for generated resources on each individualization use case and educational domain.}
    \label{tab:individualization}
\end{table}

\paragraph*{Learner Performance Evaluation}
To assess whether our ranking leads to meaningful learning gains, we conducted an experiment in which learners were randomly assigned to three groups, each receiving content ranked either first, second, or third by Voyage. Specifically, we randomly selected six topics from each of two educational domains (Python Programming and Machine Learning) to balance statistical power with feasibility under budget, time, and recruitment constraints. To emulate a realistic scenario—rapidly introducing core concepts \cite{monib2025microlearning}—we restricted materials to the brevity (level-of-detail) use case; for each of the 12 topics, the top three generated resources from our ranking were assigned to three separate groups of learners recruited via Prolific\footnote{\url{https://www.prolific.com}}, an online platform for recruiting research participants. A total of 360 participants (2 domains $\times$ 6 topics $\times$ 3 groups $\times$ 10 learners), all without prior programming or IT experience, took part. Among participants who reported demographic information, ages ranged from 18--75 years ($M=29.8$, $SD=8.6$); 42\% identified as male and 58\% as female; most reported an undergraduate or graduate degree, with a smaller subset holding a doctorate. Each participant spent ten minutes (as determined by the experts) studying their assigned resource, then completed a multiple-choice question also designed by the experts to assess the topic’s key concepts. Scores were analyzed using the \textit{Kruskal--Wallis H test}, followed by \textit{Dunn’s post-hoc tests} with \textit{Bonferroni correction}. The overall test indicated a significant difference between groups, $\chi^{2}(2, N = 360) = 15.39$, $p < .001$. As shown in \Cref{tab:learner-study-results}, learners who studied the top-ranked resource scored significantly higher than those studying the third-ranked resource, providing clear evidence that our ranking predicts learning effectiveness. It should be mentioned that the study was conducted in compliance with GDPR requirements, and all participants provided informed consent\footnote{All study data---including questions, participant demographics, and their responses---are available at \href{https://2ly.link/2F9cZ}{\texttt{https://2ly.link/2F9cZ}}}.

\begin{table}[h]
\centering
\begin{tabular}{c l c c}
\toprule
\textbf{Rank} & \textbf{Group} & \textbf{N} & \textbf{Mean Rank} \\
\midrule
1 & Group 1 (Top-ranked)   & 120 & 203.0 \\
2 & Group 2 (2nd-ranked)   & 120 & 180.5 \\
3 & Group 3 (3rd-ranked)   & 120 & 158.0 \\
\midrule[\heavyrulewidth]
\multicolumn{4}{c}{\textbf{Pairwise Comparisons (Dunn's Test, $\alpha=.017$)}} \\
\midrule
\multicolumn{2}{l}{Group 1 vs. Group 2} & .049 & Not Significant \\
\multicolumn{2}{l}{Group 1 vs. Group 3} & $<.001$ & \textbf{Significant} \\
\multicolumn{2}{l}{Group 2 vs. Group 3} & .049 & Not Significant \\
\bottomrule
\end{tabular}
\caption{Learner scores by resource rank and post-hoc pairwise comparisons. The overall Kruskal–Wallis test was significant: $\chi^2(2) = 15.39, p < .001$.}
\label{tab:learner-study-results}
\end{table}

\paragraph*{Discussion}
The learner performance study demonstrates that our ranking framework is not only aligned with expert evaluations but also translates effectively to learning gains. Learners who studied top-ranked resources significantly outperformed those using lower-ranked ones (\Cref{tab:learner-study-results}). While differences between adjacent groups (Group~1 vs. Group~2 and Group~2 vs. Group~3) did not meet the Bonferroni-corrected threshold ($p = .049 > .017$), the results showed a consistent directional pattern, with performance scores perfectly reflecting the resource rankings. This strengthens confidence that the ranking is both meaningful and practically useful. These findings are particularly important in light of our expert evaluation (\Cref{tab:individualization}), which revealed quality issues: only about half of the generated resources tailored for learners with cognitive impairments were accepted, and acceptance dropped to 45\% for those designed for learners preferring brief content. This suggests that LLMs may overlook some intended learning outcomes when generating resources for different learner needs. Notably, all LLMs produced both accepted and rejected resources, highlighting the importance of a reliable verification and ranking system. Together, the expert and learner studies demonstrate that our framework provides a scalable method for ensuring alignment with intended learning outcomes while supporting personalization and inclusiveness in educational contexts.

\section{Conclusions}\label{sec:conclusion}
The growing importance of online learning underscores the unmet need for effective and personalized learning, especially for informal and lifelong learners seeking to expand their knowledge across diverse domains. Addressing learners’ idiosyncratic needs—shaped by context, preferences, and background—requires educational providers to offer a vast array of resources. LLMs show promise for generating such resources, but a central criterion is alignment with teachers’ intended learning outcomes. In this paper, we presented two studies pertaining to a variety of educational topics. Through the studies, we
\noindent\begin{enumerate*}[label=(\arabic*)]
    \item investigated the potential of text embedding techniques for ranking educational resources according to teachers' intended learning outcomes by collecting and labeling an educational dataset,
    \item explored whether the optimal ranking solution could support teachers in validating the content quality as they generate personalized and inclusive resources, and
    \item tested whether this ranking approach translates into measurable learning benefits.
\end{enumerate*}

Our results demonstrated that the optimal text embedding model (Voyage) effectively ranked both existing and LLM-generated educational resources based on their alignment with teachers' intended learning outcomes. Additionally, the results of the learner study confirmed that higher-ranked resources led to significantly better learner performance ($\chi^{2}(2, N = 360) = 15.39$, $p < .001$), providing strong evidence that our approach not only aligns with expert evaluations but is also likely to predict learning outcomes in practice. Overall, our three-level evaluation represents a step toward a more efficient, personalized, and globally inclusive education system—one that supports teachers in offering learning materials tailored to diverse learner needs.

While our findings are encouraging, limitations should be acknowledged. We limited our scope to 53 topics and relied on \youtube{} as the source for constructing our ground truth dataset. Nevertheless, by deliberately selecting a diverse set of educational domains and leveraging \youtube{}---one of the largest platforms for informal and lifelong learning---we sought to reduce this limitation and ensure a representative evaluation. In addition, future work should explore the multilingual and multimodal aspects of generated content; however, our proposed framework is inherently language-independent, allowing for its application across different languages. Finally, the learner study was limited to short exposure per topic and single multiple-choice questions as outcome measures. Further research should replicate these findings with larger samples, more varied assessments, and longer learning interventions. We view these results as an initial demonstration rather than a conclusive evaluation; nonetheless, the convergence of expert validation and learner performance highlights the promise of embedding-based ranking as a bridge between pedagogical aims and real learner outcomes.

\bibliographystyle{unsrtnat}
\bibliography{ref}

\end{document}